\documentclass[12pt,preprint]{aastex}
\slugcomment{{\it The Astrophysical Journal, 674:329-335, 2008 February 10}}
\shorttitle{A Pre-Main-Sequence Eclipsing Binary in Orion}
\shortauthors{Cargile et al.}
\begin{document}
\title{Discovery of Par~1802 as a Low-Mass, Pre-Main-Sequence\\Eclipsing Binary in the Orion Star-Forming Region}
\author{P.~A. Cargile\altaffilmark{1},  K.~G. Stassun\altaffilmark{1}, and R.~D. Mathieu\altaffilmark{2}}
\altaffiltext{1}{Department of Physics and Astronomy, Vanderbilt University, Nashville, Tennessee 37235, USA}
\altaffiltext{2}{Department of Astronomy, University of Wisconsin-Madison, Madison, Wisconsin 53726, USA}
\begin{abstract}
We report the discovery of a pre-main-sequence (PMS), low-mass, double-lined, spectroscopic, eclipsing binary in the Orion star-forming region. We present our observations including radial velocities derived from optical high-resolution spectroscopy, and present an orbit solution that permits the determination of precise empirical masses for both components of the system. We measure that Par~1802 is composed of two equal mass (0.39$\pm$0.03, 0.40$\pm$0.03$M_{\odot}$) stars in a circular, 4.7 day orbit. There is strong evidence, such as the system exhibiting strong Li lines and a center-of-mass velocity consistent with cluster membership, that this system is a member of the Orion star-forming region and quite possibly the Orion Nebula Cluster, and therefore has an age of only a few million years. As there are currently only a few empirical mass and radius measurements for low-mass, PMS stars, this system presents an interesting test for the predictions of current theoretical models of PMS stellar evolution.
\end{abstract}
\section{Introduction}\label{Intro}
Mass is the most fundamental stellar property. The stellar masses predicted by pre-main-sequence (PMS) evolutionary models are central to our understanding of stellar formation and evolution, the timescales for circumstellar disk evolution and planet formation \citep{Haisch01, Laugnlin00}, and the initial mass function \citep{Muench02}. These models tend to systematically under-predict dynamically determined masses by 10\%-30\% \citep{Hillen04, Mathieu06}. Therefore, accurate dynamical PMS stellar mass measurements are important and necessary constraints for such models.

Low-mass stars constitute the majority of stars found in stellar populations \citep[e.g.][]{HenryMcCarthy93} and are therefore of key interest in studies of young stellar populations and in the testing and calibration of stellar formation theories. Unfortunately only five PMS empirical mass determinations are known for the mass domain below 0.5 $M_{\sun}$. One of these mass measurements \citep{Prato02} is based on distance-dependent circumstellar disk kinematics and the other four (Stassun et al.~2006; Stassun et al.~2007; Irwin et al.~in press)  are recently discovered eclipsing binary systems. Eclipsing binary stars are particularly powerful tools with which to measure these empirical stellar masses, as well as allowing other stellar properties to be determined including effective temperatures, luminosities, and radii, all in a distance-independent manner. Precise masses and radii determined from eclipsing binary systems in star-forming regions with previously known age estimates are particularly valuable because they allow constraints to be put on the age-dependent mass-radius relationship for young stars.

In this paper we report the discovery that Par~1802 is a PMS double-lined spectroscopic and eclipsing binary star system in the Orion star-forming region. The first documented photometric observations of Par~1802 came in the \citet{Parenago54} survey of the Orion Nebula Cluster (ONC). The catalog lists Par~1802 as a non-varying star with visual photographic magnitude of 15.1. Par~1802's membership to the ONC was determined in the \citet{JonesWalker88} proper-motion survey (star\# 363). In this study Par~1802 was found to have a 98\% ONC membership probability based on kinematic criteria.

More recently, a low-resolution spectral survey of the ONC \citep{Hillen97} obtained a spectral type of M2 for Par~1802. Also, \citet{Carpenter01} included Par~1802 in a time series, $JHK$ photometric survey as an extension of the 2MASS (Two Micron All Sky Survey) project. This survey produced photometric values of $J = 11.10\pm0.05$, $H = 12.50\pm0.03$, and $K_{s} = 9.97\pm0.04$~mag. We derive a $J-K_{s}$ reddening of $E(J-K)=0.29$ using a visual extinction of $A_{V}=0.72$~mag given by \citet{Hillen97} and the reddening law provided in \citet{Savage79}. Correcting for this reddening, the intrinsic color from the \citet{Carpenter01} photometry is $J-K_{s} = 0.87$. Using the \citet{Kenyon95} temperature-to-color conversion, the effective temperature of $3500K$ \citep{Hillen97} implies an intrinsic color of $J-K=0.84$. Therefore, the spectral type given in \citet{Hillen97} is consistent with the observed $JHK_{s}$ colors. The \citet{Carpenter01} survey also produced light curves that show variability in all three filters ($J H K_{s}$) and using the Stetson variability index, lists Par~1802 as a variable star.

\citet{stassun99} included Par~1802 in a 1994 December photometric monitoring survey of the ONC, but it was not flagged as a candidate eclipsing binary until it was observed spectroscopically with the WIYN multi-object spectrograph Multi-Object-Spectrograph (MOS) in 1997 January, where it showed a clear double-lined spectrum. Ex post facto examination of the $I$-band light curve revealed what appeared to be possible eclipse events, although the limited time coverage of those data coupled with the system's inherent variability (see \S\ref{PhotObs}) prevented a confident identification of eclipses. Par~1802 was subsequently monitored photometrically and spectroscopically over multiple observing seasons, which confirmed its identification as an eclipsing binary.

In \S 2 we present our photometric and spectroscopic observations of Par~1802 along with our data reduction methods. An analysis of our spectroscopic observations is detailed in \S 3, including an orbital solution from which we derive precise dynamical stellar mass measurements. In \S 4, we report our finding that Par~1802 is a young, eclipsing binary system composed of at least two low-mass stars, of approximately equal mass, in a circular orbit. In this section, we also comment on planned future work for this interesting system.

\section{Observations}\label{Obs}
\subsection{Photometric Observations}\label{PhotObs}
Par~1802 was monitored photometrically with the KPNO 0.9 m, SMARTS 1.0 m, and SMARTS 1.3 m telescopes over the period 1994 December to 2006 April. A total of 1924 $I$-band measurements were obtained on 295 different nights, with a typical cadence of 1~hr$^{-1}$, and typical photometric precision of 0.01 mag. Further observations at $VIJK$ are underway.

\begin{figure}[ht]
\epsscale{1.0}
\plotone{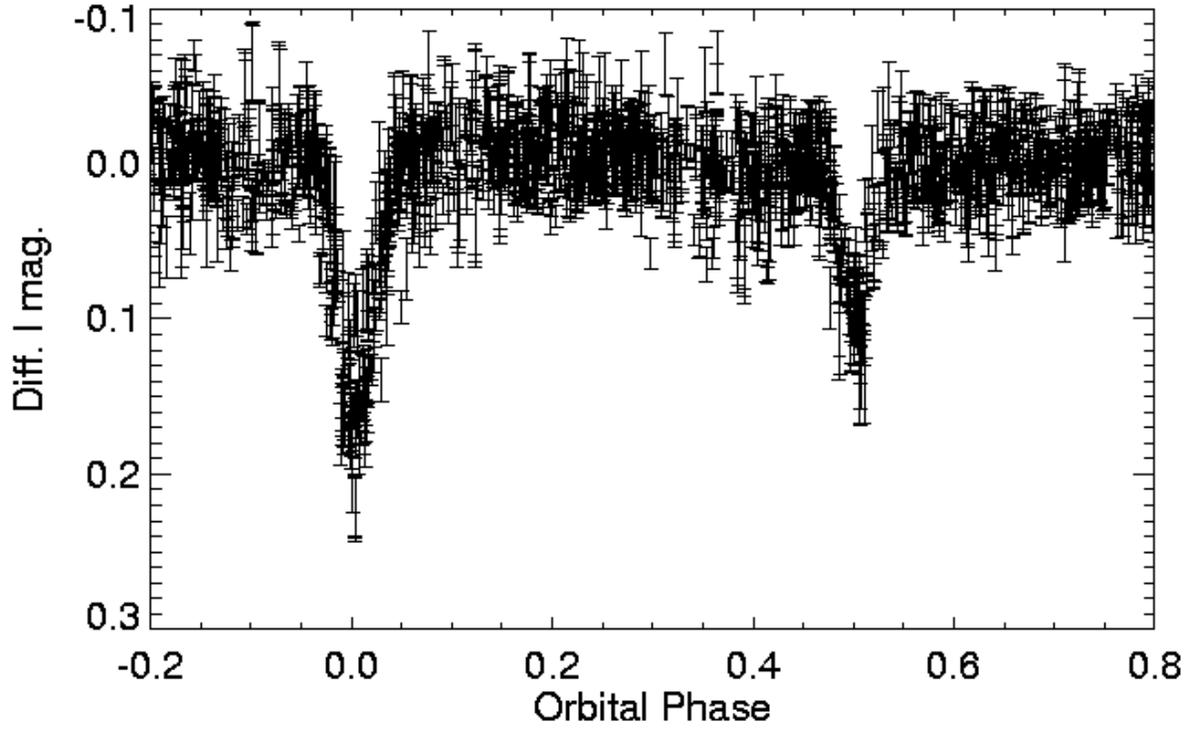}
\caption{\label{lc-fig}
The $I$-band light curve of Par~1802 folded on a period of 4.673845~days (see \S\ref{PhotObs}), and phased to the orbital time of minimum velocity separation determined from a double-lined orbit solution with $e \equiv 0$ (see \S\ref{OS}).}
\end{figure}

A period search based on the phase dispersion minimization (PDM) technique of \citet{stellingwerf78} reveals an unambiguous period of $P = 4.673845 \pm 0.000068$~days. The PDM technique is well suited to periodic variability that is highly non-sinusoidal in nature, as is the case for most eclipsing binaries. The $I$-band light curve is shown in Fig.\ \ref{lc-fig} folded on this period.

A preliminary Wilson-Devinney light-curve model gives the sum of the component radii to be $R_1 + R_2 = 3.5 \pm 0.3 {\rm R}_\odot$ which, if the components have roughly equal radii, gives $R \approx 1.75 {\rm R}_\odot$ for each component (K.~G. Stassun et al. 2008, in preparation.) These radii are much larger than the $\sim 0.5 {\rm R}_\odot$ radii expected for $0.4 {\rm M}_\odot$ main-sequence dwarfs, but are roughly consistent with the predicted radius of a $0.4 {\rm M}_\odot$ star at $\sim 1$ Myr \citep[e.g.][]{DAM97}.

These light-curve data are analyzed in detail in a follow-up study. Here we simply present the light curve for the purpose of establishing several key points. First, the light curve shows two distinct eclipses that are clearly separated in phase, as is typical of fully detached eclipsing binaries. Second, the two eclipses occur almost exactly 0.5 phase apart, suggesting a very nearly circular orbit. Third, the two eclipses are not of equal depth, indicating that the two components have unequal effective temperatures. Fourth, both eclipses are quite shallow (primary and secondary eclipse depths of $\sim 0.2$ and $\sim 0.1$ mag, respectively; Fig.\ \ref{lc-fig}), suggesting either grazing incidence or third-light dilution by a tertiary companion. Finally, there is significant scatter of $\sim 0.05$ mag peak-to-peak in the out-of-eclipse portions of the light curve, indicating the presence of intrinsic variability unrelated to eclipses. These variations have both a periodic and a stochastic component, suggesting that spots and chromospheric activity may both be present in the system.

We adopt the following ephemeris:
\begin{eqnarray}
min(I): HJD\ 2453910.40\pm0.01\ + \ 4.673845 \pm 0.000068\ E
\end{eqnarray}
where $E$ is the epoch number, and the zero point is set to the minimum of primary eclipse ($T_{0}$).

\subsection{Spectroscopic Observations}\label{SpecObs}
\subsubsection{WIYN MOS}\label{WIYNspec}
We observed Par~1802 with the WIYN\footnote{The WIYN Observatory is a joint facility of the University of Wisconsin-Madison, Indiana University, Yale University, and the National Optical Astronomy Observatories.} Multi-Object Spectrograph (MOS) on eight nights between 1997 January and 2004 December (see Table \ref{rvmeasurements-table}). We used the echelle grating centered at 6400 \AA, providing a wavelength coverage of 6250--6550 \AA\ with a resolving power of $R\approx 12000$. This spectral region contains a wealth of spectral lines and is thus well suited to radial-velocity measurements via cross-correlation. Typical exposure times were 2 hr divided into three sub-exposures for cosmic-ray rejection, yielding signal-to-noise ratio (S/N) $\sim 20$ per resolution element for most observations.  The spectra were reduced using the IRAF\footnote{IRAF is distributed by the National Optical Astronomy Observatories, which are operated by the Association of Universities for Research in Astronomy, Inc., under cooperative agreement with the National Science Foundation.\citep{Tody93}} DOHYDRA task. We flat-fielded the spectra using dome flats and wavelength-calibrated them using Th-Ar lamp spectra, each obtained with the same fiber configuration as was used for the target observations. Several fibers in each pointing were placed on blank ``sky" positions. Because these fibers were typically placed within the nebula as well, they actually represent the sum of the true sky (solar $+$ telluric) spectrum and the nebular spectrum at that location. For each pointing, the various sky fibers were median-combined to produce a single sky spectrum for that pointing, which was then subtracted from the target spectrum. Due to the small-scale variations in the nebulosity of the ONC, the median-combined sky spectrum did not completely remove the large nebular emission lines. We accounted for this in our radial velocity analysis by masking out the affected spectral regions.

Radial velocity standard stars were not observed concurrently with each target observation. However, a single high-S/N observation of the M2 standard GJ~411 was obtained on 1997 December 10 by K.\ Rhode \citep{Rhode01} using the same instrument setup. Below we use this spectrum as the radial velocity template in our cross-correlation analysis.

\subsubsection{HET HRS}\label{HETspec}
We obtained 10 high-S/N observations in queue observing mode from 2003 January to 2004 January with the High Resolution Spectrograph \citep[HRS;][]{Tull98} on the Hobby Eberly Telescope (HET).\footnote{Based on observations obtained with the Hobby-Eberly Telescope, which is a joint project of the University of Texas at Austin, the Pennsylvania State University, Stanford University, Ludwig-Maximilians-Universit\"{a}t and  Georg-August-Universit\"{a}t.} For the purpose of cosmic-ray rejection, each observation was split into two 1650 s exposures with average S/N of $\sim$50. The HET HRS setup was chosen to have a wavelength coverage from 5263 to 8915\ \AA\ and centered at 6948\ \AA\ with a resolution of R$\sim$30000 (approximately 3 pixels per resolution element). The HET HRS produces 47 orders projected onto two chips, a ``red'' chip with 17 orders and a ``blue'' chip with 30 orders. After careful examination of the resulting images from these CCDs, we found that the ``red'' chip suffered from fringing effects, causing its 17 orders to be unusable. The HET HRS also allows for a sky spectrum to be taken through a separate ``sky'' fiber, which is then dispersed and projected in between the object orders on both chips. As with the WIYN MOS, this sky spectrum generally differs compared to the Par~1802 spectrum due to variations in the nebulosity in the ONC.

We coupled each HET HRS observation of Par~1802 with an observation of a calibration Th-Ar lamp and an observation of a bright radial velocity standard star, HD~26162. Monitoring shifts in the spectra of this radial velocity standard helped to increase our understanding of the instrumental drifts associated with the HET HRS (see \S \ref{uncertOS}). 

In a previous study \citep{Stassun04} we obtained single, very high S/N ($\sim$250) spectra of late-type (K2 V, K3 V, K5 V, and K7 V) radial velocity standard stars with the same HET HRS instrumental setup as the current study. We search this archive of radial velocity standard stars to find the best template in our cross-correlation analysis (see \S \ref{RV}).

The IRAF CCDPROC and ECHELLE packages were used to process the HET HRS images, using an automated PyRAF\footnote{PyRAF is a product of the Space Telescope Science Institute, which is operated by AURA for NASA.} script that performed bias subtraction, flat-fielding, masking of bad columns, tracing of the sky and object orders, sky subtraction, and wavelength calibration. In addition, this PyRAF script was fine-tuned to handle the complexity found in HRS spectra, including highly angled orders on the CCD, flat-topped aperture profiles in the cross-dispersion direction, and having sky and object orders closely spaced on the CCD (of order a few pixels). Cosmic-ray removal was performed using the IRAF CRCOMBINE task to co-add the two 1650 s exposures of Par~1802 for a given night and reject deviant pixels.

We used the IRAF ECIDENTIFY task to derive a Th-Ar wavelength solution for each night, using a fourth-order polynomial fit in the dispersion and cross-dispersion direction. The fit involved $\sim$525 Th-Ar features identified with an rms value of 0.004\ \AA.

\section{Spectroscopic Analysis}\label{RVOrbSol}
\subsection{Radial Velocities}\label{RV}
We used the IRAF cross-correlation task FXCOR to obtain radial velocities from the WIYN MOS and HET HRS spectra of Par~1802. This method requires using a spectrum with a known heliocentric velocity as a reference velocity template \citep{TonryDavis79}.

The WIYN MOS spectrum of GJ~411 (spectral type M2) was used as the cross-correlation template for our Par~1802 WIYN MOS data. The spectra obtained during the eight nights of Par~1802 WIYN observations each produced a single cross-correlation function from which heliocentric radial velocities and their formal errors were measured using the Gaussian peak fitting tools in FXCOR. Previous experience with WIYN MOS indicates that long-term stability of the system is better than $\sim$1 km s$^{-1}$ (A.~Geller 2006, private communication). Using the \citet{TonryDavis79} r-statistic, which is a function of the cross-correlation peak width and of its height relative to the noise in the correlation function, we derived average formal uncertainties on the radial velocities of $\sim$2 and $\sim$3 km s$^{-1}$ for the primary and secondary, respectively (see Table \ref{rvmeasurements-table}).

In order to find a radial velocity template that yields the best cross-correlation with the HET HRS observations of Par~1802, we computed and compared correlation functions from each of our archived HET HRS radial velocity standard stars against a single Par~1802 epoch. The comparison led to our choice of HD~237903, our K7 V standard star, as it produced the strongest correlation peaks. To explore the possibility of Par~1802 having a later spectral type, observations and analysis of a complete set of M spectral-type templates with HET HRS are underway. 

We cross-correlated the 30 spectral orders in each Par~1802 spectrum producing 30 separate output cross-correlation functions. The cross-correlation functions were then summed for each Par~1802 epoch. By co-adding the individual order's functions the overall strength of the correlation peaks was increased, and random correlations in the spectra were minimized. We obtained a radial velocity at each epoch by measuring the centroid of a Gaussian fit to the peaks in these combined cross-correlation functions. These radial velocities are consistent, albeit with higher precision, with the velocities obtained by averaging the results from each order. The quoted uncertainties in the velocities are the statistical error in the centroid position of the Gaussian least-squares fit. We found that in some orders large, broad spectral features, such as H$\alpha$ and the \ion{Na}{2} doublet, as well as large nebular emission lines, tended to dominate the correlation function; consequently, these orders did not produce accurate centroids for radial velocity measurement. By masking out these lines during the cross-correlation process, we were able to maximize the number of correlation functions for summation. In addition, we rejected some orders due to not clearly showing both the primary and secondary correlation peaks because of other low-level structure in the correlation function. In the end, we were left with between 5 and 15 quality orders used to derive a significantly improved single combined correlation function for each HET HRS epoch of Par~1802.

\begin{figure}[ht]
\epsscale{1.0}
\plotone{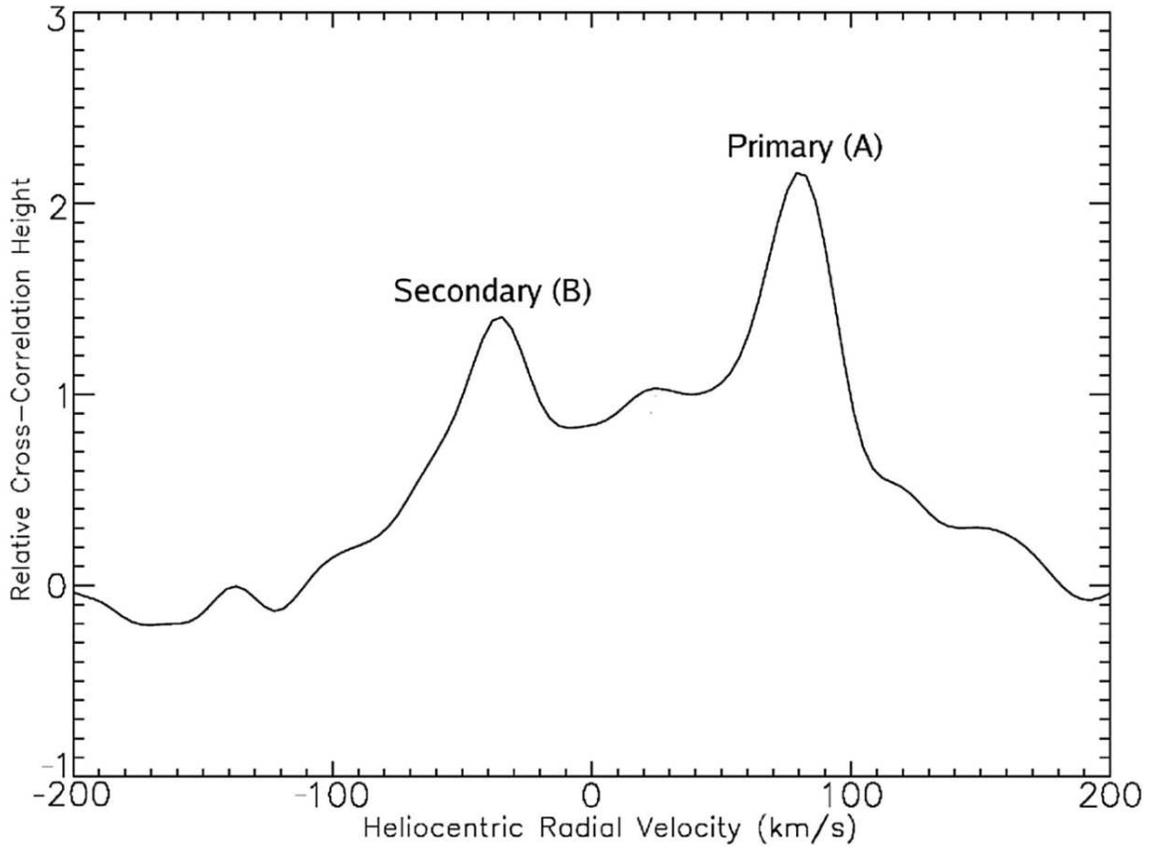}
\caption{\label{fxcor-fig}
Fourier cross-correlation function derived from the HET HRS spectrum observed at HJD$=$2452941 (orbital phase of 0.79). The function clearly shows a double-peak nature with a height ratio of $\sim$2.}
\end{figure}

Our cross-correlation functions of Par~1802 show two distinct peaks (see Fig.\ \ref{fxcor-fig}) with a height ratio of $\sim$2. By convention, we designated the component with the stronger correlation peak as the primary ($A$) and the weaker peak as the secondary ($B$). By chance a few epochs of Par~1802 spectra were taken near eclipse, resulting in a cross-correlation function with blended correlation peaks. Finding centroids of individual Gaussian functions that make up blended correlation functions is subject to many sources of uncertainty \citep{Latham96}. Therefore, radial velocities within 0.05 phase of either eclipse were not included in our modeling of the Par~1802 orbit. The remaining radial velocities have average formal measurement uncertainties of $\sim$2 km s$^{-1}$. 

\begin{figure}[ht]
\epsscale{1.0}
\plotone{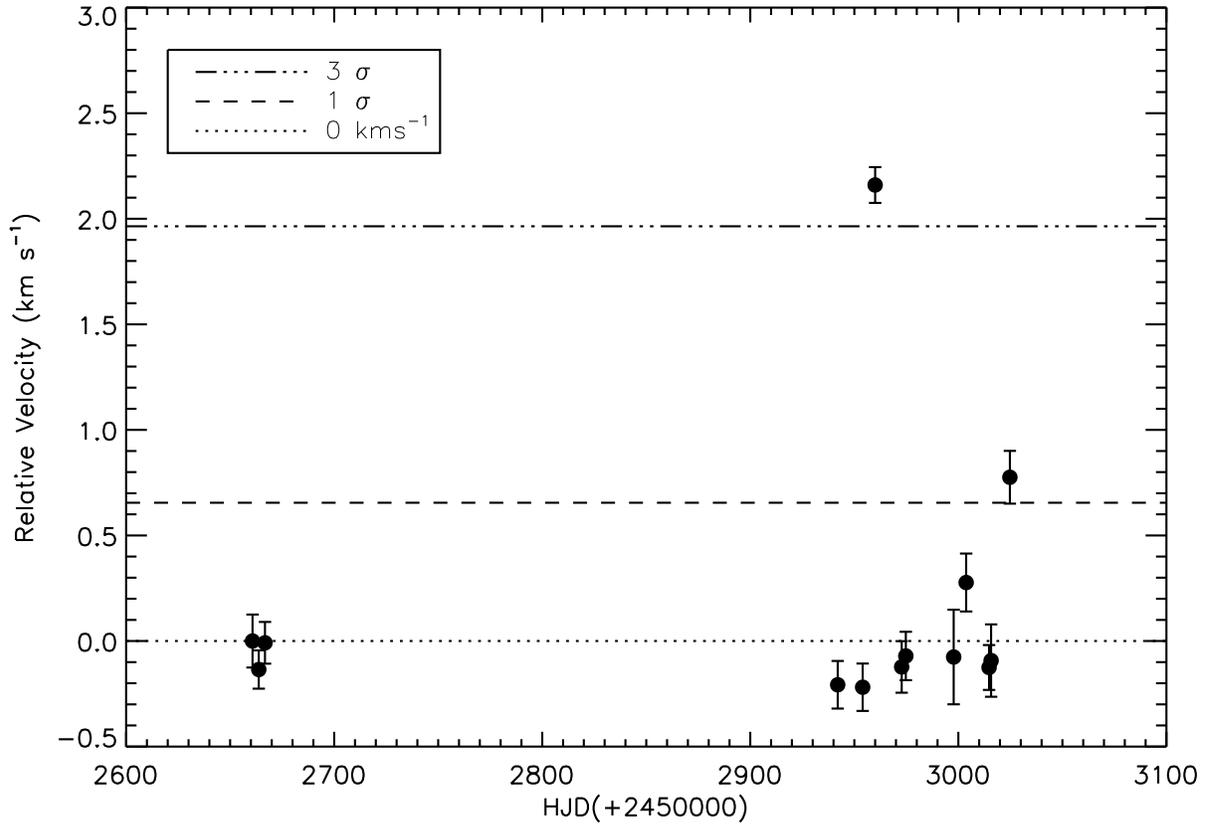}
\caption{\label{RVSTDrv-fig}
Residuals and confidence levels ({\it dashed, dotted, and solid lines}) for the radial velocity standard star, HD~26162, taken with the HET HRS on the same nights we obtained data on Par~1802. The velocities have a $\sigma$ = 0.65 km s$^{-1}$ when all points are included, and a $\sigma$ = 0.28 km s$^{-1}$ when the outlying (above 3 $\sigma$) datum is excluded.}
\end{figure}

To investigate any possible systematics in the radial velocities derived from the HET HRS spectra, we reduced the observations of the radial velocity standard star HD~26162 in the same manner as the Par~1802 data. Fig.\ \ref{RVSTDrv-fig} shows the results for this monitoring of HD~26162. The overall velocity deviations show that the HET HRS system is quite stable. Nevertheless, there was a radial velocity measurement (2.2$\pm$0.1 km s$^{-1}$ at HJD$=$2452959) obtained that deviated significantly from the mean (3.3$ \sigma$). We subtracted this epoch's instrumental shift from the Par~1802 radial velocities obtained on the same night. Except for this one outlier, the HD~26162 data reveal a high level of stability (0.28 km s$^{-1}$ rms) of the HET HRS over the 364 day time span of the HET observations and thus we opted to not subtract any offset from any other Par~1802 epoch. 

In Table \ref{rvmeasurements-table} we list in columns (1)--(4) the Julian dates, the primary and secondary radial velocities with their respective uncertainties, and the telescope used. Also listed in columns (5)--(7) are the residuals of these radial velocities relative to the orbit solution with $e \equiv 0$ (see \S\ref{OS}), and the corresponding orbital phase. We note that the residuals to the orbit solution are large compared to the uncertainties in the radial velocity measurements. A detailed analysis of those residuals and possible interpretations are provided in \S \ref{uncertOS}.

\subsection{Orbit Solution}\label{OS}
We used BINARY,\footnote{BINARY is publicly distributed by D.Gudehus at http://www.chara.gsu.edu/$\sim$gudehus/binary.html} a least-squares fitting routine, to fit an orbit solution to the radial velocity measurements from our HET HRS and WIYN MOS observations. The parameters in our orbit solution are the orbital period, $P$; the mass ratio ($M_{A}/M_{B}$), $q$; the eccentricity, $e$;  the center-of-mass velocity, $\gamma$; time of periastron passage, $T_{P}$; angle from node to periastron, $\omega$; the semi-amplitudes of the primary and secondary velocities, $K_{A}$ and $K_{B}$; the semi-major axis, $a\sin i$; and the individual masses, $M_{A} \sin^3 i$ and $M_{B} \sin^3 i$. In our modeling of the Par~1802 orbit, the orbital period was set constant to that determined photometrically, due to the precision with which we determined the system ephemeris from the light-curve data (see \S \ref{PhotObs}).

\begin{figure}[ht]
\epsscale{1.0}
\plotone{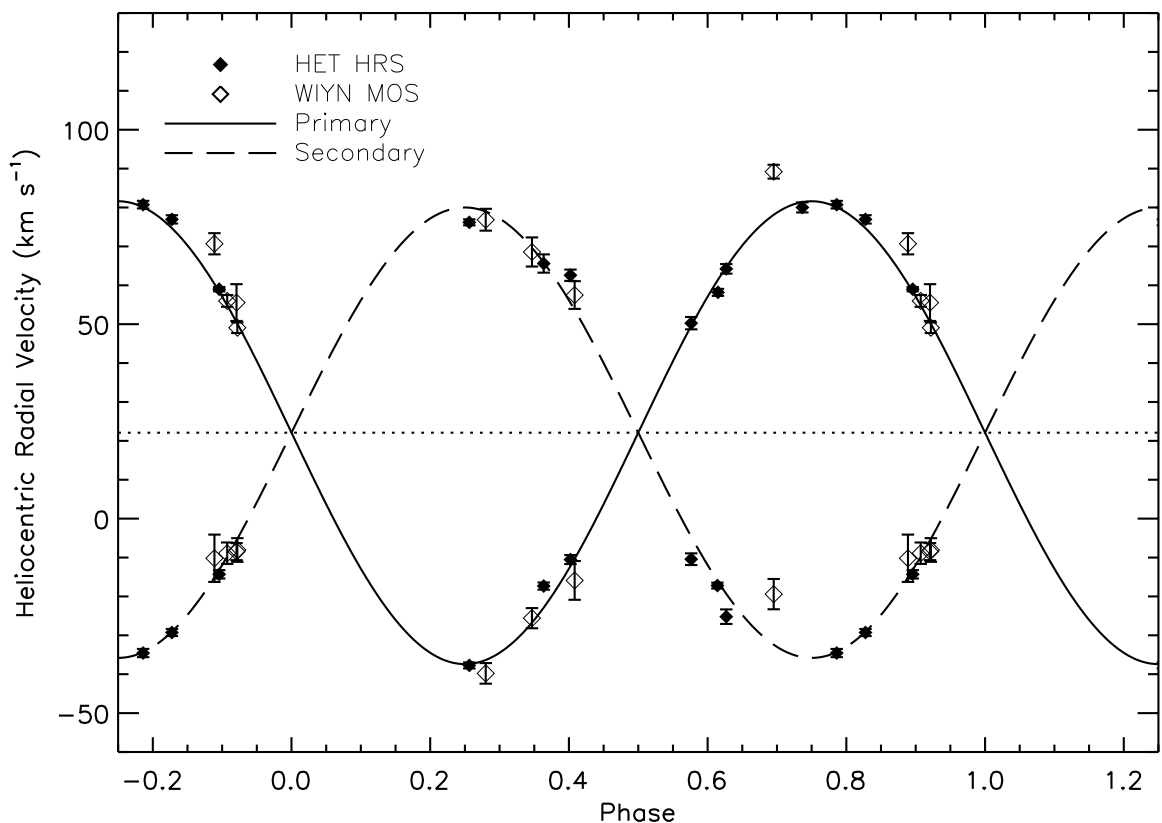}
\caption{\label{OGsol-fig}
Radial velocity orbital solution where the orbital period is held constant at the value determined by the photometric light curve, $P = 4.673845$ days, and $e \equiv 0$. Phasing is set such that the zero point corresponds to the time of primary eclipse minimum and is folded on the above period. The velocity curves show the orbital fit for the primary ({\it solid line}) and secondary ({\it dashed line}) of Par~1802. The dotted line marks the systemic radial velocity (22 km s$^{-1}$). Heliocentric radial velocity measurements are from HET HRS ({\it filled diamond}) and WIYN MOS ({\it open diamond}). We model a total of 18 epochs of observations, 10 HET HRS and 8 WIYN MOS, in this orbit solution.}
\end{figure}

We modeled an initial orbit solution in which we allowed the eccentricity to be a free parameter, with the resulting fit parameters summarized in Table \ref{orbitpars-table}, column (1). As this orbit solution is circular within the uncertainty of 0.02, we opted to fit a new orbit solution with $e \equiv 0$. The resulting parameters from this orbit solution, displayed in Fig.\ \ref{OGsol-fig}, are summarized in Table \ref{orbitpars-table}, column (2), where we give the time of primary eclipse minimum, $T_{0}$, in place of the unconstrained time of periastron passage, $T_{P}$. We note that there is no significant difference in the orbital parameters or their uncertainties ($\sigma_A$ and $\sigma_B$ in Table \ref{orbitpars-table}) between the two orbit solutions. In what follows we adopt the circular orbit solution, although we cannot formally exclude the possibility of a non-circular orbit with $e \lesssim 0.02$.

\subsection{Residuals in Orbit Solution}\label{uncertOS}
\begin{figure}[ht]
\epsscale{1.0}
\plotone{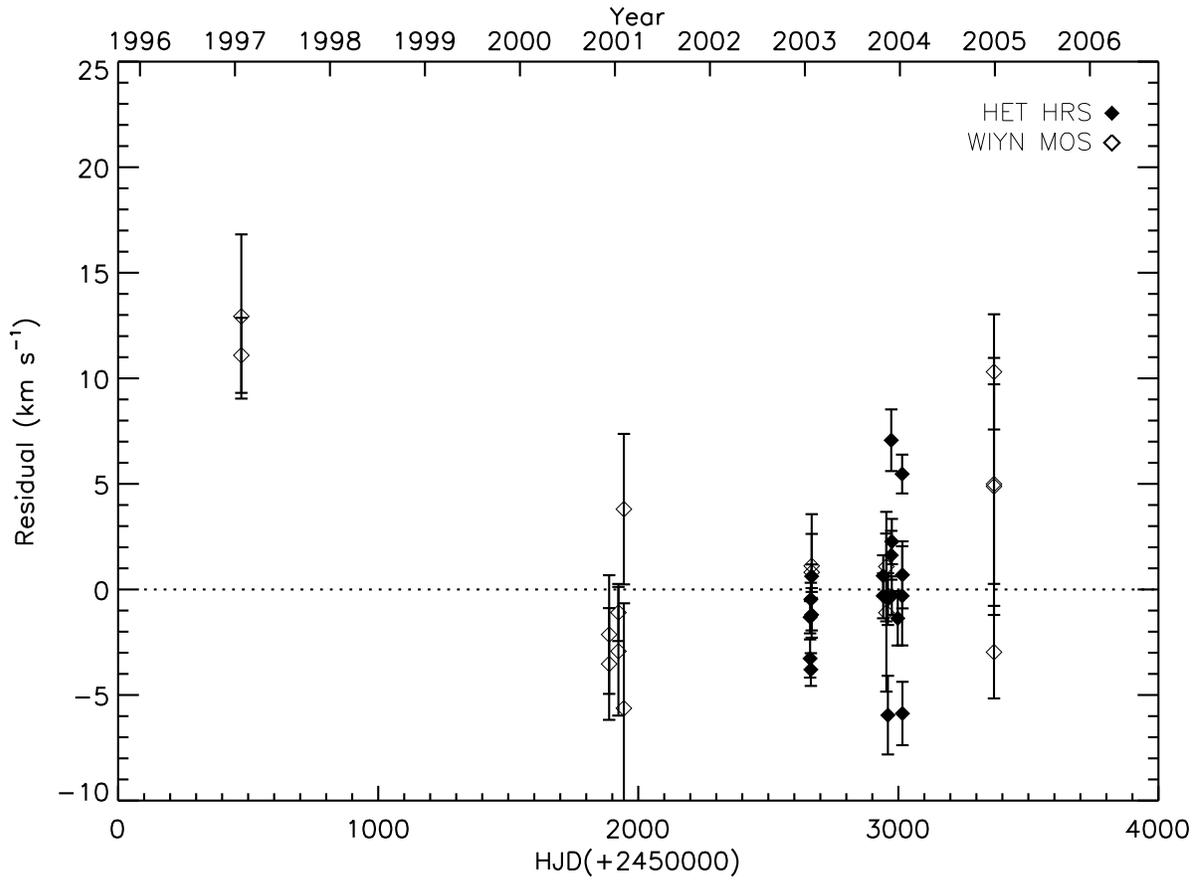}
\caption{\label{OGres-fig}
Residuals of the orbit solution (with $e \equiv 0$, see Fig.\ \ref{OGsol-fig}). Filled diamond denote radial velocity points from HET HRS and open diamond from WIYN MOS. The residuals have a standard deviation about the orbit solution of $\sigma =$ 4.5 km s$^{-1}$ or, excluding the WIYN data at HJD$=$2450473, of $\sigma =$ 3.7 km s$^{-1}$.}
\end{figure}

In Fig.\ \ref{OGres-fig} we display the residuals of our velocities about the orbit solution, shown in Fig.\ \ref{OGsol-fig}, and tabulate them in Table \ref{rvmeasurements-table}. The residuals of the WIYN radial velocity measurements at HJD$=$2450473 shows a significant offset to a large positive value. One possible interpretation is that these residuals represent a long-term systematic trend perhaps indicative of a tertiary companion orbiting the eclipsing pair with a period of $\sim$10 yr. To better estimate the periodicity of this trend, we ran a periodogram on the residuals and found a peak in the power-spectrum at 9.1 yr with a false-alarm probability of 0.2\%. Moreover, we observe what possibly is a third correlation peak in several of the cross-correlation functions, but cannot claim with certainty that this feature is always present. In any event, we examined the effect of this possible long-term trend on the fitted orbital parameters by re-running the orbit solution with the these clearly offset WIYN velocities removed, and found that the orbital parameters were unchanged within their uncertainties.

Along with the long-term systematic trend in the WIYN MOS data, we also observe significant low-level scatter in all of the radial velocity data. Excluding the outlying WIYN data at HJD$=$2450473, the radial velocities exhibit a large standard deviation (3.7 km s$^{-1}$) about the orbit solution as compared to their average measurement uncertainties ($\sim$2 km s$^{-1}$). A periodogram with the long-term systematic trend removed did not  yield any significant periodicity in the power spectrum. One possibility for this observed low-level dispersion is stellar spots. The presence of star spots can lead to asymmetries in line profiles and shifts in observed radial velocities of young, low-mass binary systems \citep[e.g.][]{Stassun04}. The photometric variability seen in our light curve (see Fig.\ \ref{lc-fig} and \S \ref{PhotObs}) also gives us strong evidence for the presence of stars spots on Par~1802. From a more detailed light-curve analysis of contemporaneous light-curve and radial velocity data we should be able to model and remove this effect from the orbit solution \citep[e.g.][]{Stassun04}.

Finally, we note that another potential source of systematic error is spectral mismatch between the radial velocity template and the components of the binary. When comparing the cross-correlation function produced by templates of different spectral types, a shift in the centroid positions of the Gaussian fit can be seen; the shift of the centroids of the cross-correlation peaks between our K5 V and K7 V templates are of order 0.2 - 0.3 km s$^{-1}$. Since Par~1802 was categorized as a M-spectral type (see \S \ref{Intro}), similar centroid shifts could be present in the measured radial velocities. Consequently, the primary and secondary velocities relative to one another will be systematically shifted. However, with orbital semi-amplitudes of $\sim$60 km s$^{-1}$, the derived mass ratio should be affected by at most $\sim$1\% by this effect.

\subsection{ONC Membership}\label{ONC}
\begin{figure}[ht]
\epsscale{1.0}
\plotone{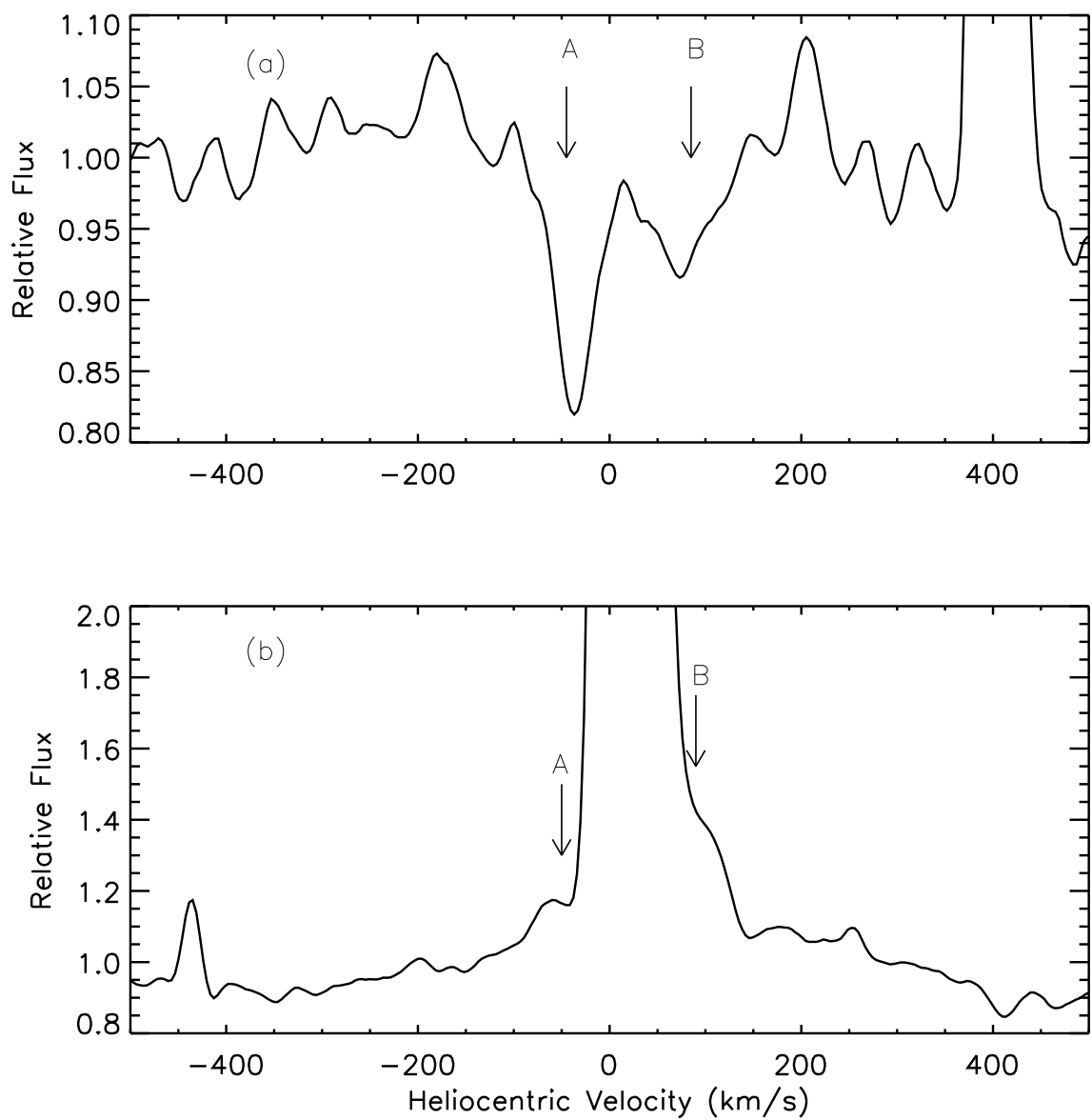}
\caption{\label{halpha-fig}
HET HRS spectra of Par~1802 taken on UT 2003 January 24 plotted vs. heliocentric radial velocity. The individual features are annotated for the primary, $A$, and secondary, $B$. ({\it a}) Par~1082 spectrum centered at the Li 6708\AA\ line. Large lines like these are features of young stars in the ONC. ({\it b}) Par~1802 spectrum centered at the H$\alpha$ line. The large emission line is due to H$\alpha$ from the Orion Nebula. The individual emission from Par~1802 $A$ and $B$ can be seen on either side of the nebular emission line. The component markers have been shifted slightly ($\sim$5 km s$^{-1}$) to avoid the large emission line.  The two components show low H$\alpha$ EQWs (a few hundred milliangstroms), which is indicative of emission most likely coming from chromospheric activity.
}
\end{figure}

In order to investigate the age of Par~1802, we inspect the system's spectra, described in \S \ref{HETspec}, for PMS signatures. We measure the Li 6708\ \AA\ line equivalent widths for the individual components of Par~1802 from a continuum-normalized, high-resolution HET HRS spectrum (HJD$=$2452664), where the two components were significantly Doppler shifted (114 km s$^{-1}$) relative to each other. We measure the lines using the spectral line deblending function in the IRAF task SPLOT. The observed equivalent widths of the two Li features are $EQW_{A}\approx210$ and $EQW_{B}\approx110 m$\AA. Correcting for the difference in continuum fluxes between the two stars (based on the relative strengths of their cross-correlation peaks; see \S\ref{RV}), we get $EQW_{A}\approx330$ and $EQW_{B}\approx300 m$\AA. Li EQW values like these are an intrinsic characteristic of PMS stars in the Orion star-forming region \citep[e.g.,][]{DolanMathieu99} observe Li EQWs above $250m$\AA\ for ONC stars). In addition, in the HJD$=$2452664 HET HRS spectrum we measure the H$\alpha$ lines of each component apart from the large nebular emission line (see Fig.\ \ref{halpha-fig}). Again using the IRAF task SPLOT, we find each component only displays a few hundred milliangstroms of H$\alpha$ emission. Classical T Tauri stars in the Orion star-forming region are found to have H$\alpha$ EQWs $>$10\ \AA\ \citep{Herbig88, Strom89}. Therefore, coupled with the small total amount of photometric variability (see \S\ref{PhotObs}), Par~1802's components do not appear to be actively accreting material from a circumstellar disk and are thus classified as weak-lined T Tauri stars.

Furthermore, our derived center-of-mass velocity for Par~1802 of $\gamma =$ 22$\pm$1 km s$^{-1}$ (see \S \ref{OS}) is comparable to the currently accepted ONC systemic cluster velocity of 25$\pm$2 km s$^{-1}$ \citep{Sicilia05}. Our kinematic membership determination is in agreement with the 98\% proper-motion ONC membership probability calculation of \citet{JonesWalker88}. 

Finally, based on a 450 pc distance to $\theta^{1}$ Ori C \citep{GenzelStutzki89}, the designated center of the ONC, the projected separation of Par~1802 from the cluster center is 1.8 pc. This places Par~1802 within 8 times the calculated core radius for the ONC \citep{HillenHart98}.

All of the preceding spectroscopic characteristics of Par~1802, plus the observed large-component radii that are similar in size to those predicted by PMS models (\S \ref{PhotObs}), lead to the conclusion that Par~1802 is a young PMS system in the Orion star-forming region and quite possibly a member of the ONC \citep[age of 1$\pm$1 Myr,][]{Hillen97}.

\section{Results and Concluding Remarks}\label{resul}
We report the discovery of a PMS eclipsing binary system, Par~1802. From our orbit solution described in \S \ref{OS}, we find the primary and secondary components are in a circular orbit with $P = 4.673845 \pm 0.000068$~days and have stellar masses of $M_{A}=$0.39$\pm$0.03 and $M_{B}=$0.40$\pm$0.03$M_{\odot}$ ($i\approxeq$ 90\degr). Par~1802 exhibits strong membership characteristics to the Orion star-forming region and quite possibly the ONC (see \S \ref{ONC}), including large Li EQWs ($\geqq 300 m$\AA) consistent with other PMS stars found in the ONC and a derived center-of-mass velocity (22$\pm$1 km s$^{-1}$) comparable to the systemic velocity of the ONC. The components of Par~1802 also have large radii ($R \approx 1.75 {\rm R}_\odot$) typical of PMS stars. We therefore estimate the age of Par~1802 to be $\sim$1 Myr. Therefore, Par~1802 $A$ and $B$ add to the recent discoveries of \citet{Stassun06} and \citet{Irwin07} to constitute the totality of empirical PMS stellar mass measurements from eclipsing binary systems below 0.5$M_{\odot}$.

We have an ongoing study of this system to further characterize the physical properties of Par~1802's components. With a simultaneous light-curve model, we will be able to determine several fundamental properties of Par~1802 $A$ and $B$ including individual radii, temperature, and $\log g$, as well as fully investigate any possible third-light contribution. In addition, to further constrain the orbit solution and investigate the long-term residual trend found in the original orbit solution, we are continuing to monitor Par~1802 with the HET HRS. We also plan a full spectral analysis to derive rotational velocity, Li abundance, and spectral classification. This future work will provide better constraints on the fundamental parameters of this important low-mass PMS binary system and thus allow precise tests of theoretical PMS evolutionary models.

\acknowledgments
We acknowledge support from the NSF Career grant AST-0349075 (PI: K.~G.~Stassun) and the NSF Grant AST 97-31302 (PI: R.~D.~Mathieu). We also gratefully acknowledge the staff at the Hobby-Eberly Telescope (HET) and D. Bell for his coordination effort for this project. The Hobby-Eberly Telescope (HET) is a joint project of the University of Texas at Austin, the Pennsylvania State University, Stanford University, Ludwig-Maximilians-Universit\"{a}t and  Georg-August-Universit\"{a}t. The HET is named in honor of its principal benefactors, William P. Hobby and Robert E. Eberly.


\clearpage

\begin{deluxetable}{ccccccc}
\tablecolumns{7}
\tablewidth{0pt}
\tabletypesize{\scriptsize}
\tablecaption{Radial Velocity Measurements\label{rvmeasurements-table}}
\tablehead{
\colhead{HJD\tablenotemark{a}}  & \colhead{RV$_{A}$}      & \colhead{RV$_{B}$}          & \colhead{Telescope\tablenotemark{b}} & \colhead{$O$-$C_{A}$\tablenotemark{c}}   & \colhead{$O$-$C_{B}$\tablenotemark{c}}   & \colhead{Phase\tablenotemark{c}}\\
\colhead{}                     & \colhead{(km s$^{-1}$)} & \colhead{(km s$^{-1}$)}      & \colhead{}                           & \colhead{(km s$^{-1}$)} & \colhead{(km s$^{-1}$)} & \colhead{}
}
\startdata
0473.6992 &  89.2 $\pm$ 1.8 & -19.4 $\pm$ 3.9 & W &  11.1 & 12.9 &  0.70\\
1887.9334 & -39.8 $\pm$ 2.6 &  76.9 $\pm$ 2.8 & W &  -3.5 & -2.1 &  0.28\\
1923.6501 &  49.1 $\pm$ 1.4 &  -8.1 $\pm$ 3.0 & W &  -1.1 & -2.9 &  0.92\\
1944.6188 & -15.9 $\pm$ 5.0 &  57.5 $\pm$ 3.6 & W &  -5.6 &  3.8 &  0.41\\
2660.6829 &  58.2 $\pm$ 0.9 & -17.2 $\pm$ 0.8 & H &  -3.3 & -1.3 &  0.61\\
2663.6809 & -37.7 $\pm$ 0.8 &  76.2 $\pm$ 0.8 & H &  -0.5 & -3.8 &  0.26\\
2666.6691 &  59.0 $\pm$ 0.6 & -14.3 $\pm$ 1.1 & H &   0.6 & -1.2 &  0.90\\
2666.7229 &  56.0 $\pm$ 1.5 &  -8.9 $\pm$ 2.8 & W &   1.1 &  0.8 &  0.91\\
2941.9133 &  80.7 $\pm$ 1.0 & -34.6 $\pm$ 1.1 & H &   0.6 & -0.3 &  0.79\\
2953.8813 & -25.6 $\pm$ 2.6 &  68.6 $\pm$ 3.7 & W &   1.1 & -1.1 &  0.35\\
2959.8645 &  66.4 $\pm$ 1.2 & -25.2 $\pm$ 1.9 & H &  -0.5 & -6.0 &  0.63\\
2972.8360 & -10.5 $\pm$ 1.2 &  62.6 $\pm$ 1.5 & H &   1.6 &  7.1 &  0.40\\
2974.8242 &  76.9 $\pm$ 1.1 & -29.3 $\pm$ 0.9 & H &   2.3 & -0.3 &  0.83\\
2997.7680 &  80.0 $\pm$ 1.3 & ...\tablenotemark{d} & H &  -1.4 &  ...\tablenotemark{d} &  0.74 \\
3014.7202 & -17.3 $\pm$ 0.9 &  65.6 $\pm$ 2.4 & H  &   5.5 & -0.3 & 0.36 \\
3015.7143 &  50.3 $\pm$ 1.6 & -10.4 $\pm$ 1.5 & H  &   0.7 & -5.9 & 0.58 \\
3367.7134 &  70.7 $\pm$ 2.7 & -10.2 $\pm$ 6.1 & W  &  10.3 &  4.9 & 0.89 \\
3367.8623 &  55.5 $\pm$ 4.7 &  -8.5 $\pm$ 2.2 & W  &   5.0 & -3.0 & 0.92 \\
\enddata
\tablenotetext{a}{Heliocentric Julian Date 2450000+}
\tablenotetext{b}{H = HET HRS and W = WIYN MOS}
\tablenotetext{c}{Derived from orbit solution with e$\equiv$0}
\tablenotetext{d}{Only able to measure primary correlation peak}
\end{deluxetable}

\clearpage

\begin{deluxetable}{c||c|c}
\tablecolumns{3}
\tablewidth{0pt}
\tabletypesize{\scriptsize}
\tablecaption{Results of Par~1802 Orbit Solution\label{orbitpars-table}}
\tablehead{
\colhead{Parameter}    & \colhead{$e$ free}  & \colhead{$e \equiv 0$}
}
\startdata
  P(d)\tablenotemark{a}         & \multicolumn{2}{c}{4.673845} \\
  $T_{0}$ (HJD)                 & \multicolumn{2}{c}{2453910.40$\pm$0.01}\\
  q                             & 0.97$\pm$0.03       & 0.97$\pm$0.03 \\
  e                             & 0.02$\pm$0.02       & 0\\
  $\gamma$ (km s$^{-1}$)        & 21.9$\pm$0.6        & 22.1$\pm$0.6\\
  $T_{P}$ (HJD)                 & 2453909.9 $\pm$0.5  & ... \\
  $\omega$ (deg)                & 54.3 $\pm$ 39.0     & ... \\
  $K_{A}$ (km s$^{-1}$)         & 60.0$\pm$1.1        & 59.5$\pm$1.0 \\
  $K_{B}$ (km s$^{-1}$)         & 58.1$\pm$1.1        & 57.9$\pm$1.1 \\
  $a \sin i$ (AU)               & 0.0507$\pm$0.0004   & 0.0511$\pm$0.0004 \\
  $M_{A} \sin^3 i\ (M_{\odot})$ & 0.40$\pm$0.03       & 0.39$\pm$0.03 \\
  $M_{B} \sin^3 i\ (M_{\odot})$ & 0.40$\pm$0.03       & 0.40$\pm$0.03 \\
  $\sigma_{A}$ (km s$^{-1}$)    & 4.40                & 4.35\\
  $\sigma_{B}$ (km s$^{-1}$)    & 5.51                & 4.79\\
\enddata
\tablenotetext{a}{Adopted from photometry.}
\end{deluxetable}

\clearpage

\clearpage

\clearpage

\clearpage

\clearpage

\clearpage

\end{document}